\documentclass[english]{article}
\usepackage[T1]{fontenc}
\usepackage[latin9]{inputenc}
\usepackage{amsmath}
\usepackage{amssymb}
\usepackage{esint}
\usepackage{color}
\makeatletter

\usepackage{babel}

\makeatother

\usepackage{babel}
\begin{document}
\begin{titlepage} \thispagestyle{empty}

\bigskip{}

\noindent \begin{center}
{\Large {An Exact Solution to the Quantized Electromagnetic Field in D-dimensional de Sitter Spacetimes}}\\

\par\end{center}

\begin{center}
\vspace{0.5cm}

\par\end{center}

\noindent \begin{center}
{G. Alencar $^{a}$%
\footnote{e-mail: geovamaciel@gmail.com %
}, I. Guedes $^{b}$, R. R. Landim $^{b}$ and R.N. Costa Filho $^{b}$}
\par\end{center}

\begin{center}
\vspace{0.5cm}
 \textit{$^{a}$Universidade Estadual do Cear\'a, Faculdade de Educa\c c\~ao, Ci\^encias e Letras do Sert\~ao Central-R. 
Epitácio Pessoa, 2554, 63.900-000  Quixad\'{a}, Cear\'{a},  Brazil. \vspace{0.2cm}
 } \textit{$^{b}$Departamento de F\'{\i}sica, Universidade Federal do Cear\'{a},
Caixa Postal 6030, Campus do Pici, 60455-760, Fortaleza, Cear\'{a}, Brazil. }
\par\end{center}

\vspace{0.3cm}

\begin{abstract}
In this work we investigate the quantum theory of light propagating in $D-$dimensional de Sitter
spacetimes. To do so, we use the method of dynamic invariants to obtain the solution of
the time-dependent Schr\"odinger equation. The quantum behavior of the electromagnetic
field in this background is analyzed. As the electromagnetism loses its conformality in
$D\neq4$, we point that there will be particle production and comoving objects will feel 
a Bunch-Davies thermal bath. This may become important in extra dimension physics and raises the intriguing possibility that precise measurements
of the Cosmic Microwave Background could 
verify the existence of extra dimensions.
\end{abstract}
\end{titlepage}

\section{Introduction}

The study of fields in a curved background has been considered along the last decades
\cite{Birrell:1982ix}. The quantization of gravity
can be carried out by using the String Theory
\cite{Polchinski:1998rq,Polchinski:1998rr,Berkovits:2000fe}. However, in cosmological
scales, gravity can be considered as a classical theory and the fields as propagating
waves in the background. Time-dependent backgrounds are used to describe many physical
systems yielding interesting results. For instance, in the study of black hole
evaporation \cite{Hawking:1974rv} and the Unruh and Casimir effects
\cite{Crispino:2007eb,Saharian:2009ii}. They are also very useful
to describe the dynamical evolution of the universe, where the production of particles in
cosmological spacetimes has been investigated
\cite{Parker:1968mv,Parker:1969au,Parker:1971pt}.

The core idea of extra dimensional models is to consider the four-dimensional universe as a hyper-surface embedded
in a multidimensional manifold. The appeal of such models is the determination of scenarios where membranes have the best chances to mimic
the standard model's characteristics. In particular, the standard model presents interesting topics to be studied such as the hierarchy
problem, and the cosmological constant problem. For instance, the Randall-Sundrum
model \cite{Randall:1999vf,Randall:1999ee} provides a possible solution to the hierarchy problem and show how gravity is trapped to a
membrane. An interesting point about this framework is that the electromagnetism loses its
conformal symmetry. Therefore, as will becomes clear later,
there can possibly be a particle production for comoving objects.

The quantum effects of a massive scalar field in the de Sitter spacetime
from a Schr\"odinger-picture point of view has been instigated in Ref.
\cite{pedrosa1}. The authors used an exact linear invariant and the Lewis and Riesenfeld
method \cite{Lewis:1968tm} to obtain the corresponding Schr\"odinger states in terms of
solutions of a second order ordinary differential equation.
By constructing Gaussian wave packet states, they calculated
the quantum dispersions, quantum correlations, and the uncertainty product for each mode of the quantized scalar field. 
Scalar fields have also been investigated in Ref. \cite{MM}, where the authors construct
the coherent states and establish the existence of squeezed states. The generalization to the case with
arbitrary dimension was given in \cite{Alencar:2011an}.

In the context of the invariant method Pedrosa {\it et al.} \cite{pedrosa2} investigated the light propagation through time-dependent dielectric linear media in the absence of free charges
and in a curved spacetime from a classical and a quantum point of view. They found that the light propagation presents a remarkable similarity in both cases. From the classical study, they
showed that the amplitude $q$ of the potential vector ( $\vec{A}$ )
is written in terms of the first- and second-order Bessel function
for a propagation through a dielectric media with $\varepsilon(t)=\varepsilon_{0}e^{\alpha t}$($\alpha>0)$. From the quantum perspective, they quantized the electromagnetic waves propagating
in a material medium with $\varepsilon(t)=\varepsilon_{0}e^{\alpha t}$, and $q$ is written in terms of harmonic functions of exponentially decaying argument for a propagation in the de Sitter spacetime 
with the metric $a(t)=e^{Ht}$, where $H$ is the Hubble constant. They wrote the solutions of the Schr\"odinger equation in terms of $\rho$, which is solution of the Milne-Pinney equation \cite{Milne,Pinney}. 
As a result they claimed that for a problem with curved spacetime the electromagnetic field quantization can also be carried out following the same steps used for the time-varying dielectric case. Moreover, 
as in the classical case, the quantum behavior of the electromagnetic field is purely oscillatory in the conformal time in the de Sitter spacetime. 

The study of de Sitter spacetime becomes important if one considers
a $\Lambda CDM$ model for the cosmos. In the present era, the universe
can approximately be described by a de Sitter spacetime, where the
matter decays with the volume and the cosmological constant is kept
constant. For $t\gg H^{-1}$, the universe tends to be fully described
by the de Sitter model. Here, we revisit the problem of quantizing the electromagnetic
field in the de Sitter space as previously considered in Ref.\cite{pedrosa2}. We present
an exact solution of the Schr\"odinger equation for arbitrary dimension
$D$. 
 
The paper is organized as follows. In Sec. II the correspondence between an
electromagnetic field
placed in a de Sitter background and a time-dependent harmonic oscillator is obtained.
In Sec. III, we use the Lewis and Riesenfeld \cite{Lewis:1968tm} method to obtain
the solutions of the related Schr\"odinger equation in arbitrary $D$. Section IV summarizes the results.

\section{Decomposition of the Vector Field}

Consider the electromagnetic field in a $D-$dimensional Friedmann-Robertson-Walker (FRW)
spacetime. The
action for the Gauge field is given by

\begin{equation}
S=-\frac{1}{4}\int d^{D}x\sqrt{-g}g^{\mu\nu}g^{\alpha\beta}F_{\mu\alpha}F_{\nu\beta},
\end{equation}
where the metric $ds^{2}=-dt^{2}+a^{2}(t)dx^{2}$ is diagonal
and $F_{\mu\nu}=\partial_{\mu}A_{\nu}-\partial_{\nu}A_{\mu}$, leading to
\begin{align}
g^{\mu\nu}g^{\alpha\beta}F_{\mu\alpha}F_{\nu\beta}=&-2\delta^{ij}a^{-2}(t)F_{0i}F_{0j}+\nonumber\\
&a^{-4}(t)\delta^{kl}\delta^{ij}F_{ki}F_{lj}.
\end{align}
Using the Gauge conditions 
\begin{equation}
A_{0}=0,\qquad \nabla\cdot\vec{A}=0
\end{equation}
we obtain 
\begin{equation}
S=\frac{1}{2}\int
d^{D}xa^{(D-1)}(a^{-2}(t)(\dot{A}_{i})^{2}-a^{-4}(t)(\partial_{i}A_{j})^{2})
\end{equation}
where the dot denotes the time derivative. Now we decompose the field as 
\begin{equation}
\vec{A}(\vec{x},t)=\sum_{\lambda=1}^{D-2}\int\frac{d^{D-1}k}{(2\pi)^{D-1}}\hat{\epsilon}^{(\lambda)}_k\left[u_{k}^{(\lambda)}(t)e^{i\vec{k}\cdot\vec{x}}+u_{k}^{(\lambda)*}(t)e^{-i\vec{k}\cdot\vec{x}}\right]
\end{equation}
where $\hat{\epsilon}$ represents a unit vector satisfying
$\vec{k}\cdot\hat{\epsilon}^{(\lambda)}=0$
due to the Gauge condition and $\hat{\epsilon}^{(\lambda')}_{k}\cdot\hat{\epsilon}^{(\lambda)}_{k}=\delta^{\lambda'\lambda}$,
and $\lambda=1,...,D-2$ stands for the $D-2$ polarizations. By separating $u_{k}^{(\lambda)}$ into real and imaginary parts 
\begin{equation}
u_{k}^{(\lambda)}=\frac{1}{\sqrt{2}}(q_{k1\lambda}+iq^{2}_{k2\lambda})
\end{equation}
the action reads 
\begin{align}
S=&\frac{1}{2}\int
dt\int\frac{d^{D-1}k}{(2\pi)^{3}}\times\nonumber\\
&\sum_{\lambda,i=1}^{D-2}\left[a(t)^{(D-3)}
\dot{q}_{ki\lambda} ^{2}-k^{2}a^{(D-5)}(t)q_{ki\lambda}^{2}\right],\label{acao}
\end{align}
 where $i=1,2$ labels the real and imaginary parts of $u_{k}^{(\lambda)}$.
From Eq.(\ref{acao}) we obtain the Hamiltonian for the electromagnetic
field 
\begin{equation}
H_{\lambda ik}=\frac{1}{2}a^{-(D-3)}(t)p_{\lambda ik}^{2}+\frac{1}{2}k^{2}a^{(D-5)}(t)q_{\lambda ik}^{2},
\end{equation}
where 
\begin{equation}
p_{\lambda ik}=\frac{\partial L}{\partial \dot{q}_{\lambda ik}}=a(t)^{(D-3)}\dot{q}_{\lambda ik},
\end{equation}
with p being the conjugate momentum. The classical equation of motion
for the $qth$ mode reads 
\begin{equation}
\ddot{q}_{\lambda ik}+(D-3)\frac{\dot{a}}{a}\dot{q}_{\lambda ik}+\frac{k^2}{a^2}q_{\lambda ik}^{2}=0.\label{classicaleq}
\end{equation}

From the above Hamiltonian and equation of motion we can see that each mode is decoupled. Therefore they can be 
treated independently.
Next, consider the harmonic oscillator with time-dependent
mass and frequency given by the Hamiltonian 
\begin{equation}
H(t)=\frac{p^{2}}{2m(t)}+\frac{1}{2}m(t)\omega^{2}(t)q^{2},
\label{hamiltoniano}
\end{equation}
where $[q,p]=i\hbar$. The equation of motion reads 
\begin{equation}
\ddot{q}+\frac{\dot{m}(t)}{m(t)}\dot{q}+\omega^{2}(t)q=0,
\label{HO}
\end{equation}
which is similar to Eq. (\ref{classicaleq}) if one considers that each mode of the electromagnetic field
corresponds to the time-dependent harmonic oscillator with $m(t)=a^{D-3}(t)$
and $\omega=ka^{-1}(t)$. Therefore our problem is reduced to find a solution
to the time dependent Harmonic Oscillator. For this we will use the 
Emarkov approach.

\section{Quantization of Electromagnetic Fields in the de Sitter Spacetime}
\subsection{The Emarkov Approach}
Consider a time-dependent harmonic oscillator described by Eq. (\ref{hamiltoniano}).
It is well known that an invariant for Eq. (\ref{hamiltoniano}) is
given by \cite{Carinena}

\begin{equation}
I=\frac{1}{2}\left[\left(\frac{q}{\rho}\right)^{2}+(\rho p-m\dot{\rho q})^{2}\right]\label{invariantedef}
\end{equation}
where $q(t)$ satisfies Eq.(\ref{HO}) and $\rho(t)$ satisfies the
generalized Milne-Pinney equation \cite{Milne,Pinney}

\begin{equation}
\ddot{\rho}+\gamma(t)\dot{\rho}+\omega^{2}(t)\rho=\frac{1}{m^{2}(t)\rho^{3}}\label{MP}
\end{equation}
with $\gamma (t)=\dot{m}(t)/m(t)$. The invariant $I(t)$ satisfies the equation

\begin{equation}
\frac{dI}{dt}=\frac{\partial I}{\partial t}+\frac{1}{i\hbar}[I,H]=0
\end{equation}
and can be considered hermitian if we choose only the real solutions
of Eq. (\ref{MP}). Its eingenfunctions, $\phi_{n}(q,t)$, are assumed
to form a complete orthonormal set with time-independent discrete eigenvalues,
$\lambda_{n}=(n+\frac{1}{2})\hbar$. Thus

\begin{equation}
I\phi_{n}(q,t)=\lambda_{n}\phi_{n}(q,t)\label{invariante}
\end{equation}
with $\left\langle \phi_{n},\phi_{n'}\right\rangle =\delta_{nn'}$.
Taking he Schr\"odinger equation (SE)

\begin{equation}
i\hbar\frac{\partial\psi(q,t)}{\partial t}=H(t)\psi(q,t)\label{SE}
\end{equation}
where $H(t)$ is given by Eq. (\ref{hamiltoniano}) with $p=-i\hbar\frac{\partial}{\partial q}$,
Lewis and Riesenfeld \cite{Lewis:1968tm} showed that the solution $\psi_{n}(q,t)$
of the SE (see Eq.(\ref{SE})) is related to the functions $\phi_{n}(q,t)$
by 

\begin{equation}
\psi_{n}(q,t)=e^{i\theta_{n}(t)}\phi_{n}(q,t)
\end{equation}
where the phase functions $\theta_{n}(t)$ satisfy the equation

\begin{equation}
\hbar\frac{d\theta_{n}(t)}{dt}=\left\langle \phi_{n}(q,t)\left|i\hbar\frac{\partial}{\partial t}-H(t)\right|\phi_{n}(q,t)\right\rangle .
\end{equation}
The general solution of the SE may be written as

\begin{equation}
\psi (q,t)=\sum_{n}c_{n}e^{i\theta_{n}(t)}\phi_{n}(q,t)
\end{equation}
where $c_{n}$ are time-independent coefficients. Now, using an unitary transformation and following the steps drawn in Ref. \cite{Carinena} we find

\begin{align}\label{psi}
\psi_{n}(q,t)=&e^{i\theta_{n}(t)}\left(\frac{1}{\pi^{1/2}\hbar^{1/2}n!2^{n}\rho}\right)^{1/2}\times\nonumber\\
&\exp\left\lbrace\frac{im(t)}{2\hbar}\left[\frac{\dot{\rho}}{\rho}+\frac{i}{m(t)\rho^{2}(t)}\right]q^{2}\right\rbrace\times \\
&H_{n}\left(\frac{1}{\sqrt{\hbar}}\frac{q}{\rho}\right)\nonumber
\end{align}
where 

\begin{equation}
\theta_{n}(t)=-(n+\frac{1}{2})\int_{t_{0}}^{t}\frac{1}{m(t')\rho^{2}}dt'\label{theta},
\end{equation}
and $H_{n}$ is the Hermite
polynomial of order $n$. Therefore the quantization of the time dependent HO depends on finding a solution
to the associated MP equation (\ref{MP}) to be included in Eq. (\ref{psi}). A solution to this non-linear equation is given by a 
non-linear combination of the solutions to the linear case \cite{Prince}. Observe that the linear version of the MP equation is identical to our
classical equation of motion (\ref{HO}). Therefore, if we can
find solutions of (\ref{HO}) we can find the desired solution to the problem. 

Now we get back to the quantization of the Electromagnetic Field. For $m(t)=a^{D-3}(t)$ and $\omega(t)=ka^{-1}(t)$ the auxiliary
Equation (\ref{MP}) reads 
\begin{equation}
\ddot{\rho}+(D-3)\frac{\dot{a}}{a}\dot{\rho}+k^{2}a^{-2}(t)\rho=\frac{a^{-2(D-3)}(t)}{\rho^{3}}.\label{MPa}
\end{equation}
We should point that $\rho$ does not depends on $\lambda,i$ and this dependence is solely from $q_{\lambda i k}$. As commented above, to find the solution for Eq. (\ref{MPa}), we will first look for solution of the classical equation (\ref{classicaleq}). For this we will follow Ref. \cite{Finelli:1999dk}. We must
consider the conformal time $dt=a(t)d\eta$, and 
define a new function by $q_{\lambda ik}=\Omega\bar{q}_{\lambda ik}$. With this, Eq. (\ref{classicaleq}) reads
 
\begin{eqnarray}
&&\bar{q}_{\lambda ik}''+\left(2a\frac{\dot{\Omega}}{\Omega}-\dot{a}+(D-3)\dot{a}\right)\bar{q}_{\lambda ik}'\nonumber \\ 
&&+\left(k^{2}+a^2\frac{\ddot{\Omega}}{\Omega}+(D-3)a\dot{a}\frac{\dot{\Omega}}{\Omega})\right)\bar{q}_{\lambda ik}=0\label{classicaleqD}
\end{eqnarray}
where in the above equation the prime and the dot means a derivative with respect to $\eta$ and $t$ respectively. by choosing $\Omega=a^{-\frac{D-3}{2}}$ we find
\begin{eqnarray}
&&\bar{q}_{\lambda ik}''-\dot{a}\bar{q}_{\lambda ik}'+[k^{2}-\frac{(D-3)}{2}a\ddot{a}\nonumber \\
&&+\frac{(D-3)(D-5)}{4}\dot{a}^2]\bar{q}_{\lambda ik}=0
\end{eqnarray}

We would like to point that the case $D=3$ cancel many terms and we get a
simplified equation depending on the choice of $a$. This can become important for condensed matter systems. Consider now the de Sitter spacetime, $a=e^{Ht}$. With $dt=a(t)d\eta$ we get the relations
\begin{equation}
\eta=-\frac{e^{-Ht}}{H}=\frac{1}{Ha(t)},\;\dot{a}=-\frac{1}{\eta},\;  \ddot{a}=-\frac{H}{\eta} 
\end{equation}
and we obtain
\begin{equation}
\bar{q}_{\lambda ik}''+\frac{1}{\eta}\bar{q}_{\lambda ik}'+(k^{2}-\frac{(D-3)^2}{4\eta^2})\bar{q}_{\lambda ik}=0. 
\end{equation}

Finally we get
\begin{equation}
\left[\frac{d^2}{d(k\eta)^2}+\frac{1}{(k\eta)}\frac{d}{d(k\eta)}+(1-\frac{\nu^2}{(k\eta)^2})\right]\bar{q}_{\lambda ik}=0. 
\end{equation}
where $\nu^2=\frac{(D-3)^2}{4}$. This is a Bessel equation with solutions given by $J_\nu(k|\eta|)$ and $N_\nu(k|\eta|)$. Using now our previous redefinition
of $q$ we find for the solutions to the classical equation (\ref{classicaleqD}) 
$$
\begin{cases}
a^{\frac{-(D-3)}{2}}J_{\nu}(k|\eta|)\\a^\frac{-(D-3)}{2}N_{\nu}(k|\eta|) 
\end{cases}
$$
Now we can come back to the Milne-Pinney equation (\ref{MPa}). It is known that its solution can be found from the above solutions of the classical equation of motion as \cite{Finelli:1999dk}
\small
\begin{eqnarray}
 \rho=\frac{a^{\frac{-(D-3)}{2}}H2^{\frac{1}{2}}}{\pi}\left[AJ_\nu^2+BN_\nu^2+(AB-\frac{\pi^2}{4H^2})^{\frac{1}{2}}J_\nu N_\mu\right]^\frac{1}{2}
\end{eqnarray}
\normalsize
where $A$ and $B$ are real constants. The fixing of these constants is related to the choice of our vacuum. This is due to the fact that the construction
of particle states and the choice of the vacuum is not unique in curved spaces as the one used here. This is important since the production
of particles can be inferred only after we choose some vacuum to compare with our physical solution. A natural choice is the Bunch-Davies vacuum, which is
the adiabatic vacuum at early times. For this we can fix $A=B=\pi/2H$ and $\rho$ reads, for $a=1/H\eta$   
\begin{eqnarray}
 \rho=(|H|\eta)^{\frac{(D-3)}{2}}\sqrt{\frac{H}{\pi}} \left[J_\nu^2+N_\nu^2\right]^\frac{1}{2}.\label{solution}
\end{eqnarray}
which is the general solution for $\rho$ with arbitrary $D$. The solution for the electromagnetic field now can be found by substituting this
in Eq. (\ref{psi}). The fact $\rho$ is time dependent for $D \neq 4$ has important consequences that we discuss in the conclusions. 
\subsection{The $D=4$ Conformal Time Approach}
For $D=4$, Eq. (\ref{solution}) yields $\rho=1/k^2$. In this case, from Eqs. (\ref{theta}) and (\ref{psi}) with $m(t)=a(t)$, the solution to the SE  is given by
\begin{align}
\psi_{n}(q,t)=& \left(\frac{k}{\pi\hbar n!^{2}2^{n}}\right)^{\frac{1}{4}}e^{-ik\left(n+\frac{1}{2}\right)\int_{t_0}^{t}a^{-1}(t)}\times\nonumber\\
&\exp\left(-\frac{kq^{2}}{2\hbar}\right)H_{n}\left(\sqrt{\frac{k}{\hbar}}q\right).\label{conformaltime}
\end{align}

However, only for this case, there is a simpler way of obtaining the above solution. For doing so, we use the fact that the electromagnetic field is conformal in $D=4$. Consider the conformal time
$dt=a(t)d\eta$ as before. The metric is changed to $ds^{2}=-a^2(t)(d\eta^{2}+dx^{2})$ 
and we get that in the conformal time we have a free electromagnetic field, whose quantization is well known, namely
\begin{align}
\psi_{n}(q,t)=& \left(\frac{k}{\pi\hbar n!^{2}2^{n}}\right)^{\frac{1}{4}}e^{-ik\left(n+\frac{1}{2}\right)\left(\eta-\eta_0)\right)}\times\nonumber\\
&\exp\left(-\frac{kq^{2}}{2\hbar}\right)H_{n}\left(\sqrt{\frac{k}{\hbar}}q\right).
\end{align}
By performing the inverse transformation we obtain Eq. (\ref{conformaltime}).
This kind of reasoning has been used to point that the comoving referential does not led to particle production. 
It should be observed that any nontrivial solution of Eq. (\ref{MPa}) will give a different solution
of Eq. (\ref{psi}). This can be obtained if we consider cases with $D\neq4$.

\section{Concluding remarks}

In this paper we used the Lewis and Riesenfeld method to obtain the
time-dependent Schr\"odinger states emerging from the quantization of
the electromagnetic field in the $D$-dimensional de Sitter spacetime. It is well-known
that a challenge in obtaining the exact solution (see Eq. (\ref{psi}))
for the SE with $H$ given in Eq. (\ref{hamiltoniano}), is the solution
of the Milne-Pinney (in terms of the universal scale factor of universe
$a(t)$) equation (\ref{MPa}). For $D=4$ a solution can be found trivially in the conformal time. 
This kind of reasoning has been used to point that the comoving referential do not leds to a particle production. However if we consider the space
with dimensions different than four, the electromagnetic field looses its conformality. Therefore, as said in the introduction, this can have important 
consequences for condensed matter or extra dimension physics.

For $D\neq4$ we can easily verify that $\rho=constant$ is not a solution of equation above. In the case of a de Sitter spacetime, 
we found the general solution  and as we expected, the electromagnetic field has more interesting solutions. For time-dependent solutions, 
there will be particle production and the comoving referential in this space will feel a thermal bath. This is important for planar system in condensed matter. A very intriguing
consequence of this is for extra dimension physics. This has gained a lot of attention due to superstring theory \cite{Polchinski:1998rq,Polchinski:1998rr} and in brane-model universes. 
In this model our universe is conceived as a brane in a five dimensional space. A de Sitter
space time would therefore imply a particle production and a thermal bath for the comoving referentials in this enlarged space. Therefore, at least in principle,
this could contribute with an effective temperature in the membrane and raises the intriguing possibility that
extra dimensions could be found by precise measure in the Cosmic Microwave Background.

At last, we would like to point out that the procedure described can
be used to trace the present properties of the quantum electromagnetic
field back to the recombination era in an arbitrary $D$-dimensional universe. This would be a much more interesting
phenomenological result. 

\section{Acknowledgments}

The authors would like to thank Jailson S. Alcaniz for useful discussions. We would also like to thank: The Goethe-Institut Berlin for the hospitality. We acknowledge the financial
support provided by Funda\c c\~ao Cearense de Apoio ao Desenvolvimento Cient\'\i fico e Tecnol\'ogico (FUNCAP), the Conselho Nacional de Desenvolvimento Cient\'\i fico e Tecnol\'ogico (CNPq) and FUNCAP/CNPq/PRONEX.

\end{document}